\documentclass[runningheads,english]{llncs}[2022/01/12]
\usepackage{amsmath}
\usepackage{upquote}
\usepackage[main=english]{babel}
\usepackage[T1]{fontenc}
\usepackage{graphicx}
\usepackage{hyperref}
\hypersetup{
  hidelinks,
  colorlinks=true,
  allcolors=blue,
  pdfstartview=Fit,
  breaklinks=true
}
\usepackage[frozencache,newfloat]{minted}
\newcommand{\labelline}[1]{\label[line]{#1}\hypertarget{#1}{}}
\newcommand{\refline}[1]{\hyperlink{#1}{\FancyVerbLineautorefname~\ref*{#1}}}
\usepackage[autostyle=true]{csquotes}
\usepackage[group-minimum-digits=4,per-mode=fraction]{siunitx}
\usepackage[all]{hypcap}
\usepackage[capitalise,nameinlink]{cleveref}
\usepackage{tikz}
\usetikzlibrary{shapes,positioning,shadows,arrows.meta,calc}
\begin{document}
\title{On the usefulness of linear types for correct nonce use enforcement during compile time}
\titlerunning{Enforcing unique nonces with linear types}
\author{Richard Ostertág\orcidID{0000-0002-6560-1515}}
\authorrunning{R. Ostertág}
\institute{Faculty of Mathematics, Physics and Informatics,\\
Comenius University, Bratislava, Slovakia\\
\email{ostertag@dcs.fmph.uniba.sk}}
\maketitle
\begin{abstract}
Cryptographic algorithms and protocols often need unique random numbers as parameters (e.g. nonces). Failure to satisfy this requirement lead to vulnerable implementation and can result in security breach. We show how linear types and static type checking can be used to enforce the correct generation of a~new unique random number for each function invocation.

\keywords{Secure coding \and Nonce \and Linear types \and Rust.}
\end{abstract}
\section{Motivation}

The security of various cryptographic constructions relies on unique or even unpredictable values. Examples are nonces in cryptographic protocols, initialization vectors in modes of symmetric encryption, salts in password-based key derivation functions etc. These values are often generated as a random numbers of prescribed length.

Programmers, which are not experts in cryptography, may believe that it is not strictly necessary to generate a new random number every time. Programmers can be lazy and provide some numeric constant instead of a new random number for each use. After all, the cryptographic construction will “correctly”\footnote{Depending on the construction it can for example still correctly encrypt and decrypt messages.} work even with this fixed numeric constant. However, if the no-reuse principle is not followed, it can lead to a serious security vulnerability in the resulting application (which is not visible at first glance). Well known example is forbidden attack for AES-GCM \cite{Nonce-Disrespecting}, but e.g. see also \cite{Joux}.

We show how to implement a cryptographic library that would allow the compiler to detect incorrect (i.e. repeated) use of such one-time random numbers at compile time. We will divide this task into two parts:
\begin{enumerate}
\item In the first part, we ensure that the function expecting a random number gets as an argument a random number generated by an “approved” method.

E.g. a true random number generated by a special hardware device and not just software generated pseudorandom number or we can enforce usage of any chosen specific software implementation.

For this first part, we will utilise abstract data types with a hidden data constructor.

\item In the second part, we will ensure that once the generated random number is used, it cannot be reused for the second time.

For the second part, we will use linear types. We will illustrate on the Rust programming language, but the idea can be used in any programming language with linear types.

\end{enumerate}

\section{Abstract data types}

An abstract data type (ADT) is defined by its behaviour (e.g. operations like insert or delete). However, the implementation details are hidden from its users. The implementers have the flexibility to use data stractures internally or even make changes to their approach in the future. As long as the external behaviour (interface) remains unchanged, all existing code that uses this ADT can function without requiring any updates to adapt to any modifications made to the internal implementation.

ADTs are commonly used and supported in many standard programming languages, for example C++, Java, Pascal. ADTs are usually realised as modules or objects concealing internal implementation and exposing only the public interface. For example, if we want to realise the stack (FIFO data structure) as ADT, we will provide public functions like \mintinline{rust}{push}, \mintinline{rust}{pop}, \dots and type \mintinline{rust}{Stack} for variables holding values of this ADT. But the important aspect is, that we do not provide the client with any information on how the stack is internally implemented. It may be a linked list or an array or something totally different. We also do not provide any external means for creating a new stack (because external users do not know the internal details of the \mintinline{rust}{Stack} type). The only possibility to create a new stack is to call some function from the module, which returns a new \mintinline{rust}{Stack} value (or create a new instance if objects are used instead of modules).

ADT are useful for constraining access and preventing invalid states. By creating the stack as ADT, the implementer of the module can maintain strict control over its representation. A client has no way to accidentally (or maliciously) alter any of the stack representation invariants.

We can use this technique to create a \mintinline{rust}{nonce} module in Rust with \mintinline{rust}{Nonce} abstract data type (see \cref{lst:mod}). We have created a public struct type \mintinline{rust}{Nonce} with a private random value of type \mintinline{rust}{u128}. The client can not directly create structs with any private fields. In this case for example it is invalid to write \mintinline{rust}{let nonce = Nonce { val: 42 }}. The only way for the client to create a nonce is to call a public constructor method \mintinline{rust}{let mut nonce = nonce::Nonce::new()}. Because the client needs to call the \mintinline{rust}{new} method we can guarantee that on \refline{line:random} we, as implementers, choose the right system function to generate a new random number (e.g. we may use hardware RNG).

\begin{listing}[htbp]
  \begin{minted}[linenos=true,escapeinside=||]{rust}
mod nonce {
    // A public struct with a private random value of type u128
    pub struct Nonce {
        val: u128,
    }

    impl Nonce {
        pub fn new() -> Nonce {     // A public constructor method
            use rand::prelude::*;
            Nonce { val: random() }      |\labelline{line:random}|
        }

        pub fn get(&self) -> u128 { // A public getter method
            self.val
        }
    }
}
  \end{minted}
  \caption{Implementation of nonce module in Rust}
  \label{lst:mod}
\end{listing}

While abstract types are a powerful means of controlling the structure and creation of data, they are not sufficient to limit the ordering and number of uses of values and functions. As another example, we can mention e.g. files. There is no (static) way to prevent a file from being read after it has been closed. Also, we cannot stop a client from closing a file twice or forgetting to close a file at all. In our case, there is no static way to stop the client from using one nonce value multiple times just with ADT. But this can be enforced in programming languages with linear types.

\subsection{Linear types}
Before presenting our proposed solution (using linear types), we want to quickly recapitulate what linear types are \cite{Pierce04} and how they are implemented in the well-known Rust programming language \cite{Rust}.

Linear types are a special case of substructural type systems, which are particularly useful for constraining interfaces that provide access to system resources such as files, locks and as we will show, we can constrain random number reuse. Substructural type systems augment standard type abstraction mechanisms with the ability to control the number and order of uses of a data structure or operation, which is exactly what we need.

\subsection{Structural Properties}
Lets discuss three basic \emph{structural} properties. The first property, \emph{exchange}, indicates that the order in which we write down variables in the context is irrelevant. A corollary of exchange is that if we can type check a term with the context $\Gamma$, then we can type check that term with any permutation of the variables in $\Gamma$. 

\begin{equation}\tag{Exchange}
\frac
{\overbrace{\Gamma_1, x\!:\!\tau_x, y\!:\!\tau_y, \Gamma_2}^{\textrm{context}} \vdash e\!:\!\tau}
{\underbrace{\Gamma_1, y\!:\!\tau_y, x\!:\!\tau_x, \Gamma_2}_{\textrm{permutated context}} \vdash e\!:\!\tau}
\end{equation}

The second property, \emph{weakening}, indicates that adding extra, unneeded assumptions to the context, does not prevent a term from type checking. 

\begin{equation}\tag{Weakening}
\frac
{\Gamma \vdash e\!:\!\tau}
{\Gamma, \underbrace{x\!:\!\tau_x}_{\makebox[0pt]{$\scriptstyle\textrm{unneeded assumption}$}} \vdash e\!:\!\tau}
\end{equation}

Finally, the third property, \emph{contraction}, states that if we can type check a term using two identical assumptions ($x_2\!:\!\tau_{x_1}$ and $x_3\!:\!\tau_{x_1}$) then we can check the same term using a~single assumption.

\begin{equation}\tag{Contraction}
\frac
{\Gamma, x_2\!:\!\tau_x, x_3\!:\!\tau_x \vdash e\!:\!\tau}
{\Gamma, x_1\!:\!\tau_{x_1} \vdash [x_2 \mapsto x_1, x_3 \mapsto x_1]e\!:\!\tau}
\end{equation}

\subsection{Substructural Type Systems}

A \emph{substructural type system} is any type system that is designed so that one or more of the structural properties do not hold \cite{Pierce04}. Different substructural type systems arise when different properties are withheld.

\begin{description}
\item[Linear type systems] ensure that every variable is used exactly once by allowing exchange but not weakening or contraction.
\item[Affine type systems] ensure that every variable is used at most once by allowing exchange and weakening, but not contraction.
\item[Relevant type systems] ensure that every variable is used at least once by allowing exchange and contraction, but not weakening.
\item[Ordered type systems] ensure that every variable is used exactly once and in the order in which it is introduced. They do not allow any of the structural properties.
\end{description}

The picture below can serve as a mnemonic for the relationship between these systems. The system at the bottom of the diagram (the ordered type system) admits no structural properties. As we proceed upwards in the diagram, we add structural properties: E stands for exchange; W stands for weakening; and C stands for contraction. It might be possible to define type systems containing other combinations of structural properties, such as contraction only or weakening only, but so far researchers have not found applications for such combinations \cite{Pierce04}. Consequently, they are excluded them from the diagram.

\begin{figure}[h!]
\centering
\begin{tikzpicture}[auto, to/.style={thick,->,>=stealth}]
  \node (unrestricted)                             {unrestricted (E,W,C $\Rightarrow$ structural)};
  \node (affine)   [below left  = of unrestricted] {affine (E,W)};
  \node (relevant) [below right = of unrestricted] {relevant (E,C)};
  \node (linear)   [below = 2.5cm of unrestricted] {\textbf{linear (E)}};
  \node (ordered)  [below       = of linear]       {ordered (none)};
  \path[to] (ordered)  edge node[swap] {+E} (linear);
  \path[to] (linear)   edge node {+W} (affine) edge node[swap] {+C} (relevant);
  \path[to] (affine)   edge node {+C} (unrestricted);
  \path[to] (relevant) edge node[swap] {+W} (unrestricted);
\end{tikzpicture}
\caption{Relationship between linear and other substructural type systems.} \label{fig1}
\end{figure}
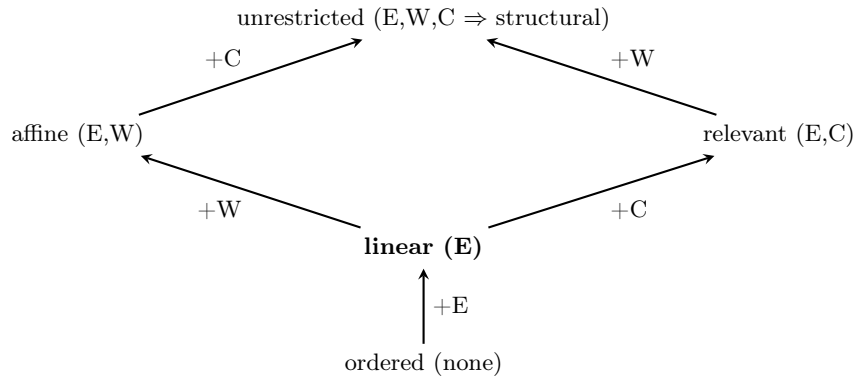

\subsection{Rust}

Ownership is Rust’s most unique feature. It enables Rust to make memory safety guarantees without needing a garbage collector. The feature is straightforward to explain. In Rust, the memory is managed through a system of ownership with a set of rules, that the compiler checks at compile time. None of the ownership features slow down the program while it is running (unlike garbage collection).

\subsubsection{Ownership rules}

\begin{itemize}
\item Each value in Rust has a variable that is called its \emph{owner}.
\item There can be only one owner at a time.
\item When the owner goes out of scope, the value will be dropped (memory will be deallocated).
\end{itemize}

\noindent We will demonstrate these rules on \cref{lst:ownership}. On \refline{line:create} we create a string and assign its value into variable \mintinline{rust}{s1}. This variable is now the only owner of the string. Then on \refline{line:move} we move value from variable \mintinline{rust}{s1} to new owner -- variable \mintinline{rust}{s2}. Now \mintinline{rust}{s2} is the only owner of the string value. That is the reason, why we can not use variable \mintinline{rust}{s1} on \refline{line:borrow} to borrow the string value to \mintinline{rust}{println!} function. But we could use \mintinline{rust}{s2} for this. When \mintinline{rust}{s2} comes out of scope the string value can be deallocated from memory.
\vspace{-1em}

\begin{listing}[htbp]
  \begin{minted}[linenos=true,escapeinside=||]{rust}
fn borrowing() {
    let s1 = String::from("Hello");      |\labelline{line:create}|
// move ^^ occurs because `String` does not implement the `Copy` trait
    let s2 = s1; // value moved here     |\labelline{line:move}|
    println!("{}, world!", s1); // value borrowed here after move    |\labelline{line:borrow}|
}

  \end{minted}
  \vspace{-1.5em}
  \caption{Example of ownership rules}
  \label{lst:ownership}
\end{listing}

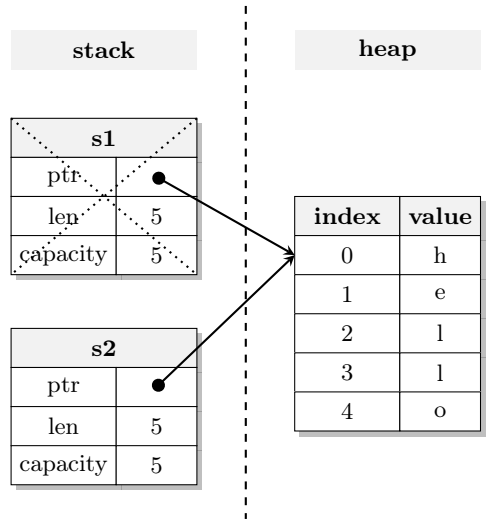
\begin{figure}[h!]
\centering
\begin{tikzpicture}[
node distance=0cm,outer sep=0pt,fill=white,text centered,minimum height=1.6em,anchor=north west,
box/.style={fill=black!5,minimum width=7.6em},
boxT/.style={draw,fill=black!5,minimum width=7.6em,drop shadow},
boxL/.style={draw,fill,        minimum width=4.3em,drop shadow},
boxR/.style={draw,fill,        minimum width=3.3em,drop shadow}
]
	\node (stack) [box] {\textbf{stack}};
	\node (heap)  [box,right = 4em of stack] {\textbf{heap}};

    \node (s1)  [boxT,below = 2em of stack]{\textbf{s1}};
    \node (pl1) [boxL] at (s1.south west)  {ptr};      \node (pr1) [boxR] at (pl1.north east) {};
    \node (ll1) [boxL] at (pl1.south west) {len};      \node (lr1) [boxR] at (ll1.north east) {5};
    \node (cl1) [boxL] at (ll1.south west) {capacity}; \node (cr1) [boxR] at (cl1.north east) {5};
   
    \node (s2)  [boxT,below = 7em of s1]   {\textbf{s2}};
    \node (pl2) [boxL] at (s2.south west)  {ptr};      \node (pr2) [boxR] at (pl2.north east) {};
    \node (ll2) [boxL] at (pl2.south west) {len};      \node (lr2) [boxR] at (ll2.north east) {5};
    \node (cl2) [boxL] at (ll2.south west) {capacity}; \node (cr2) [boxR] at (cl2.north east) {5};
    
    \node (idx) [boxL,fill=black!5,right = 4em of lr1] {\textbf{index}}; \node (val) [boxR,fill=black!5] at (idx.north east) {\textbf{value}};
    \node (i0)  [boxL] at (idx.south west) {0};  \node (v0)  [boxR] at (i0.north east)  {h};
    \node (i1)  [boxL] at (i0.south west)  {1};  \node (v1)  [boxR] at (i1.north east)  {e};
    \node (i2)  [boxL] at (i1.south west)  {2};  \node (v2)  [boxR] at (i2.north east)  {l};
    \node (i3)  [boxL] at (i2.south west)  {3};  \node (v3)  [boxR] at (i3.north east)  {l};
    \node (i4)  [boxL] at (i3.south west)  {4};  \node (v4)  [boxR] at (i4.north east)  {o};

    \draw [Circle-stealth,thick,shorten <=-1.5pt] (pr1.center) -- (i0.west);
    \draw [Circle-stealth,thick,shorten <=-1.5pt] (pr2.center) -- (i0.west);

	\draw [dotted,thick] (s1.north west) -- (cr1.south east);
	\draw [dotted,thick] (s1.north east) -- (cl1.south west);
	
	\draw [dashed,thick] ($(stack.north west)!.5!(heap.north east)+(0,1em)$) --
	               ($(stack.north west)!.5!(heap.north east)-(0,20em)$);
\end{tikzpicture}
\caption{Memory representation of variables \mintinline{rust}{s1} and \mintinline{rust}{s2}} \label{fig2}
\end{figure}

This is illustrated in \cref{fig2}. After assigning to \mintinline{rust}{s2} the value from \mintinline{rust}{s1}, variable \mintinline{rust}{s2} points to the same memory on the heap, but \mintinline{rust}{s1} can not be used for dereferencing anymore. This is used primarily for memory management without the need for a garbage collector or explicit deallocation. We will use these ownership rules for constraining nonce usage.

\section{The solution}

The solution in Rust is syntactically  very simple because it is well aligned with Rust syntax. Usually, when functions in Rust take arguments, they are passed as references (with \mintinline{rust}{&} before variable name). This way value is not moved to the parameter from the local variable (it is just borrowed). But we prevent this in \mintinline{rust}{Nonce} type, because we do not implement \mintinline{rust}{Copy} trait. Traits are similar to interfaces in other languages. To read more about traits see for example \cite{Traits}.

\begin{listing}[htbp]
  \begin{minted}[linenos=true,escapeinside=||]{rust}
fn need_new_random_u128_every_time(nonce: nonce::Nonce) {
    let _tmp = nonce.get();
    println!("Nonce param value: {}", nonce.get());
    println!("Nonce param value: {}", *nonce);      |\labelline{line:star}|
}
  \end{minted}
  \caption{Example of function with nonce as argument}
  \label{lst:fn}
\end{listing}

On \cref{lst:fn} we implement function \mintinline{rust}{need_new_random_u128_every_time} to demonstrate function signature for functions that require new random value for every call. The body of the function is not significant, but we demonstrate, that the nonce value can be used repeatedly inside library implementation, which is often needed. We also implement  \mintinline{rust}{Deref} trait, so \mintinline{rust}{*} can be used on \refline{line:star} instead  of longer \mintinline{rust}{nonce.get()} from the line above.

When function \mintinline{rust}{need_new_random_u128_every_time} is called, then value ownership is moved from the local variable to the argument and thus local variable can not be used anymore. As an example, if in \cref{lst:main} we comment out \refline{line:comment}, we will get compile time error “value used here after move” on the next line.

\begin{listing}[htbp]
  \begin{minted}[linenos=true,escapeinside=||]{rust}
fn main() {
    // Structs with private fields
    // can be created only using public constructors
    let mut nonce = nonce::Nonce::new();
    need_new_random_u128_every_time(nonce);

    nonce = nonce::Nonce::new();                    |\labelline{line:comment}|
    need_new_random_u128_every_time(nonce);

    need_new_random_u128_every_time(nonce::Nonce::new());
}
  \end{minted}
  \caption{Example of nonce usage}
  \label{lst:main}
\end{listing}

\section{Conclusion}

We have demonstrated how to use ADT and linear types in Rust for enforcing the freshness of nonces for library function calls. In Rust, the syntax is very straightforward. This solution can be implemented also in other languages with linear types, like Haskell, which experimentally supports linear types from version 9.0.1. But syntax, in this case, is not so clear as in Rust.

\subsubsection{Acknowledgements}
This publication is the result of support under the Operational Program Integrated Infrastructure for the project: Advancing University Capacity and Competence in Research, Development and Innovation (ACCORD), co-financed by the European Regional Development Fund.

\bibliographystyle{splncs04}
\bibliography{nonce}

\end{document}